\documentclass[aps,pra,floatfix,amsmath,preprint,showpacs,11pt]{revtex4}
\usepackage{graphicx}
\usepackage{dsfont}
\begin{document}

\title{Using the Hadamard and related transforms for 
simplifying the spectrum of the quantum baker's map}
\author{Arul Lakshminarayan, N. Meenakshisundaram}
\affiliation{Department of Physics\\ Indian Institute of Technology Madras\\
Chennai, 600036, India.}

\preprint{IITM/PH/TH/2006/6}

\begin{abstract}
We rationalize the somewhat surprising efficacy of the Hadamard transform in 
simplifying the eigenstates of the quantum baker's map, a paradigmatic model of
quantum chaos. This allows us to construct closely related, but new, transforms that do
significantly better, thus nearly solving for many states of the 
quantum baker's map. These transforms, which combine the standard Fourier and Hadamard
transforms in an interesting manner, are constructed from eigenvectors of the shift
permutation operator that are also simultaneous eigenvectors of bit-flip (parity) and 
possess bit-reversal (time-reversal) symmetry.
\end{abstract}
\maketitle


\newcommand{\newc}{\newcommand}
\newc{\beq}{\begin{equation}}
\newc{\eeq}{\end{equation}}
\newc{\kt}{\rangle}
\newc{\br}{\langle}
\newc{\beqa}{\begin{eqnarray}}
\newc{\eeqa}{\end{eqnarray}}
\newc{\pr}{\prime}
\newc{\longra}{\longrightarrow}
\newc{\ot}{\otimes}
\newc{\rarrow}{\rightarrow}
\newc{\h}{\hat}
\newc{\bom}{\boldmath}
\newc{\btd}{\bigtriangledown}
\newc{\al}{\alpha}
\newc{\be}{\beta}
\newc{\ld}{\lambda}
\newc{\sg}{\sigma}
\newc{\p}{\psi}
\newc{\eps}{\epsilon}
\newc{\om}{\omega}
\newc{\mb}{\mbox}
\newc{\tm}{\times}
\newc{\hu}{\hat{u}}
\newc{\hv}{\hat{v}}
\newc{\hk}{\hat{K}}
\newc{\ra}{\rightarrow}
\newc{\non}{\nonumber}
\newc{\ul}{\underline}
\newc{\hs}{\hspace}
\newc{\longla}{\longleftarrow}
\newc{\ts}{\textstyle}
\newc{\f}{\frac}
\newc{\df}{\dfrac}
\newc{\ovl}{\overline}
\newc{\bc}{\begin{center}}
\newc{\ec}{\end{center}}
\newc{\dg}{\dagger}
\newc{\prh}{\mbox{PR}_H}
\newc{\prq}{\mbox{PR}_q}

\section{Introduction}
The eigenstates of quantized chaotic systems continue to be studied vigorously and
present several challenges. Exactly or nearly exactly solvable models are desirable 
in this setting, but are few and far in between, especially for systems that 
display generic behaviour. We had pointed out earlier \cite{meen} that the Walsh-Hadamard (WH)
 transform (or simply the Hadamard transform) \cite{Schroeder} simplified the eigenstates of the
  quantum baker's map \cite{Qbmap}, a paradigmatic model of quantum chaos, considerably, the so-called
 ``Thue-Morse'' state being a particularly good, even spectacular, example of this.
  Further, we showed that an exactly solvable permutation operator, viewed as
the quantization of cyclic shifts, was useful in constructing a relevant basis \cite{aruljphys}. 
We connect these two observations here and explicitly point out the
 rationale for the emergence of the Hadamard transform
in the context of the quantum baker's map, and therefore are able to 
go beyond by constructing a transform based on the Hadamard that simplifies
the eigenstates of the quantum baker's map much more significantly. It is quite likely that
the new transform finds use in broader contexts since it combines the widely
 used discrete Fourier transform and the Hadamard transform in a novel manner.

We briefly review the relevant background, leaving some details that maybe found in the
references. The baker's map is a textbook example of a simple fully chaotic system.
 The classical baker's map \cite{LL}, $T$, is the area preserving
 transformation of the unit square $[0,1)\times [0,1)$ onto itself, which takes a
phase space point $(q,p)$ to $(q',p')$ where 
\beq
\left(q', p'\right) 
= \left\{ \begin{array}{cl}(2q,\, p/2) & \mbox{if} \;\;0\le q<1/2 \\
(2q-1,\,(p+1)/2)& \mbox{ if} \;\; 1/2\le q<1.\end{array} \right.
\eeq
The repeated action of $T$ on the unit square leaves the phase
space completely mixed, this is well known to be a fully chaotic system that 
is Bernoulli. The area-preserving property makes this map a model of chaotic
 two-degree of freedom Hamiltonian systems, and the Lyapunov exponent is
$\log(2)$ per iteration.  

The classical baker's map is exactly solvable in many ways, including an 
explicit prescription for finding periodic orbits of any period, which follows
on using the binary representation for the phase space variables. The action
of $T$ as a complete left-shift becomes clear in this representation.
 Due to its simplicity, and due to the fact that there is nothing but chaos
 (it is not a perturbation of an integrable system), its quantization due to
 Balazs and Voros \cite{Qbmap} has been used extensively in studies of quantum
  chaos and semiclassical methods. It has also been experimentally implemented
   recently using NMR \cite{NMR}.
   
 The ``usual'' quantum baker's map, in the position representation is:
\beq \label{bvsbak} B=G_N^{-1}\left( 
\begin{array}{cc}G_{N/2} &0\\0 & 
G_{N/2} \end{array}\right), \eeq %
where 
\begin{displaymath}
(G_{N})_{nm}= \df{1}{\sqrt{N}}\,\exp\left(\df{-2 \pi i}{N} 
(n+1/2)(m+1/2)\right).
\end{displaymath}

We require that
 $N$ be an even integer; Saraceno \cite{Qbmap} imposed anti-periodic boundary
  conditions that we use. The Hilbert space is finite dimensional, the dimensionality
   $N$ being the scaled inverse Planck constant $(N=1/h)$, where we have used that the 
phase-space area is unity. The position and momentum states are denoted as 
$|q_n\kt$ and $|p_m\kt$,where $m,n=0,\cdots,N-1$ and the transformation function 
$\br q_n|p_m \kt $ between these bases is the finite Fourier transform $(G_N)_{nm}
$ given above.
 
 The choice of anti-periodic boundary conditions fully preserves parity
  symmetry, here called $R$, which is such that $R|q_{n}\kt = |q_{N-n-1}\kt$. 
Classically this is the symmetry $(q \rightarrow 1-q,\, p \rightarrow 1-p)$.
  Time-reversal symmetry is also present and implies in the context of
  the quantum baker's map that an overall phase can be chosen such that the
momentum and position representations are complex conjugates: $G_N
\phi=\phi^{*}$, if $\phi$ is an eigenstate in the position basis. 
$B$ is an unitary matrix, whose repeated application is the quantum version
 of the full left-shift of classical chaos. Unlike the quantum cat map, where
 many analytical results concerning the spectrum is known \cite{Cat}, this is not the case with
 $B$. However, the quantum map $B$ is more generic than the quantum cat map, and hence it's
 spectrum, especially the eigenstates, is of considerable interest.

We had shown that a simple exactly solvable shift operator $S$,
 acts as a good zero-th order operator for the quantum baker's map, although 
 there is no classical integrable ``zero-th order''
 system for the baker's map \cite{aruljphys}. Here we also showed how one can construct
  a quantum baker's map $B_S$ that is very closely allied to $B$ above, using $S$, thereby explaining this
 closeness. It may therefore be expected that the eigenstates of $S$ form a basis
  in which the eigenstates of the quantum baker's map  appear simple. Thus we seek to
  diagonalize the shift operator and use this as a basis to expand the eigenstates of the
  quantum baker's map.

   However the spectrum of $S$ is typically highly degenerate, especially when
    $N$ is a power of $2$, therefore we have multiple choices for the
  eigenstates. In \cite{aruljphys} we had used $N$ that lead to the smallest degeneracy
   possible. Specifically those $N$ whose order modulo 2 is the largest possible value,
    $N-2$, have a nondegenerate spectrum save for two states with unit eigenvalues.
    The case when $N$ is a power of 2 is however very interesting, as the eigenfunctions
    are well simplified using the Walsh-Hadamard transform, and there are connections to
    automatic sequences such as the Thue-Morse sequence \cite{meen}.
     Therefore it is natural that we seek to understand this case from the point of view
      that uses the shift operator essentially. To do this we make use of the symmetries
       of the shift operator  especially the parity and time-reversal to ``reduce'' the
        eigenstates. The use of quantum symmetries is of course natural,
	and we note that in the case of quantum cat maps \cite{Cat} the complete use of all quantum
	symmetries results in exactly solvable states that are ergodic and these have been 
	called Hecke eigenfunctions \cite{rudnick}.

\section{Eigenfunctions of the Shift operator}

We will henceforth denote the position eigenkets $|q_n\kt$ simply as $|n\kt$ and
use this as a basis unless otherwise stated.     
The shift operator $S$, by definition, acts on the position basis as
 \beq
 S|n\kt = \left\{ \begin {array}{ll} |2n\kt &0\le n <N/2\\ 
|2n-N+1\kt & N/2 \le n \le N-1. \end{array}\right.
\eeq
 The shift operator $S$ is a generalization of what was proposed as
  the quantum baker's map by Penrose \cite{Penrose} for the case when $N=2^K$, $K$ integer.
  In this case, which is of sole concern in this paper,
   the Hilbert space is isomorphic to that of $K$ qubits, or two-level systems.
    Let $n=a_{K-1}a_{K-2}\cdots a_0$ be the binary expansion of $n$, so that 
\beq
|n \kt = |a_{K-1}\kt \otimes |a_{K-2} \kt \otimes \cdots |a_{0}\kt,
\eeq 
where now the ``qubit'' states $|0\kt$ and $|1\kt$ are orthonormal basis states. In the 
standard representation 
\beq 
|0\kt \doteq \left( \begin{array}{c}1\\0\end{array} \right), \;\; 
|1\kt \doteq \left( \begin{array}{c}0\\1\end{array} \right).
\eeq
Of course the state $|0\kt $ is not to be confused with the state $|n=0\kt$ which is 
actually $\otimes ^K |0 \kt$. In the following we will not work with individual
qubits states as such for this confusion to arise.

The action of the shift operator is then transparent: 
\beq 
S|n\kt =  |a_{K-2}\kt \otimes |a_{K-3}\kt \cdots |a_0\kt \otimes  |a_{K-1}\kt.
\eeq
Thus the quantum operator $S$ embodies the left-shift action, by cyclically shifting the
states from one qubit to its left "neighbour". It however re-injects the
most significant bit at the least significant position, due to periodic 
boundary conditions, which ultimately naturally leads to periodicity. 
It is also helpful to think of $S$ as acting on the space $\{0,1\}^K$ consisting
of binary strings of length $K$, which we denote generically by $\sg$: $S(a_{K-1}a_{K-2}
\cdots a_{0})=a_{K-2}a_{K-3} \cdots a_{0} a_{K-1}$. 

The parity operator $R$ introduced earlier, can be written
as a pure $K$-fold tensor product on this space, so that it can be thought of as local
action on the individual qubits.
\beq
R=\otimes^K {\cal R}, \;\;\; \mbox{where}\;\;\;  {\cal R}= 
\left( \begin{array}{cc}0&1\\1&0 \end{array} \right).
\eeq
Thus the action on individual qubits is that of a flip. When considered as action
on binary strings, $R(a_{K-1}a_{K-2}\cdots a_{0})=\ovl{a}_{K-1}\ovl{a}_{K-2}\cdots 
\ovl{a}_{0}$, where $\ovl{0}=1$ and $\ovl{1}=0$ are bit-flips.
It is easy to see that $S$ commutes with $R$. While there are uncountably many operators 
that commute with $S$ (any operator of the form $\otimes^K {\cal A}$, where ${\cal A}$
is a 1-qubit operator, will commute with $S$), $R$ also commutes with the usual 
quantum baker's map $B$, and is thus an important symmetry for constructing a basis that 
is close to that of the eigenfunctions of the quantum baker's map. The classical limit of
 the unitary operator $S$ has been discussed earlier \cite{aruljphys}.
 It can also be thought of as quantizing a multivalued mapping that
  is hyperbolic \cite{aruljphys}, similar interpretations have been proposed
  for toy models of open bakers in \cite{Nonnen}. However we continue to use it here only in so far as
   it enables us to understand the eigenstates of $B$. 

It is particularly simple to diagonalize $S$ in the case when $N=2^K$. Let $d$ be a 
divisor of $K$ (including $1$ and $K$). Let $\sg=s_1 s_2 \cdots s_{d}$
 be a {\it primitive} binary string of length $d$, representing 
 strings consisting of all its cyclic shifts. For example there are 12 primitive strings of 
 length 4, consisting of three cycles. We represent these three cycles with the strings
 $0001,0011,$ and $0111$, chosen for convenience, to be the smallest when the cycle elements
 are evaluated as numbers. Let $\overline{\sg}$ denote the string $\sg$ repeated $K/d$ times
 and its value evaluated as a binary representation be $k$. 
  
  Then one set of eigenstates of $S$ constructed from these cycles are:
 \beq 
  |\tilde{\phi}_l^k\kt = \f{1}{\sqrt{d}} \sum_{m=0}^{d-1}  e^{-2 \pi i lm/d}\, S^m |k \kt,\;\; 0\le l \le d-1
\eeq
and the corresponding eigenvalue is $e^{2\pi i l/d}$ \cite{aruljphys}. If there are $p(d)$ such primitive 
representative strings of length of $d$, then $\sum_{d|K} d \,p(d)= 2^K$, where using the
Mobi\"us inversion formula, $p(n)=\sum_{k|n}\mu\left(n/k\right)\, 2^k/n$,
 ($\mu(n)$ is the Mobi\"us function, $\mu(n)=0$
if $n$ has a repeated prime in its prime factorization, otherwise it is $(-1)^r$, where 
$r$ is the number of primes in its factorization, and $\mu(1)=1$.)
 Thus $|\tilde{\phi}_l^k \kt$ form a complete and orthonormal set. Clearly there is
degeneracy and this set is not unique. There is also freedom in the choice of the
 representative string $\sg$, we will choose this to be the smallest integer when
 treated as a binary expansion. The eigenvectors arranged in columns of a matrix constitute
 an unitary transform which consists of direct sums of 
 $p(d)$ Discrete Fourier Transforms (DFTs) of dimension $d$ each. It may be written as:
 \beq
 T_f=\bigoplus_{d|K} \bigoplus_{p(d)} \, F_d
 \eeq
 where $F_d$ is a DFT of dimension $d$. For example if $K=3$, the divisors are only
  $1$ and $3$, there are two subspaces of dimension $1$ and
 two of dimension $3$ corresponding to the cycles $\ovl{0},\, \ovl{1},\ovl{001},\ovl{011}$. 
 If the basis is arranged in the order $000$, $001$, $010$, $100$, $011$, $110$, $101$, $111$, we have chosen
 the first member of the cycle to be the smallest, the others are got by consecutive 
 left-shifts, $T_f=F_1\bigoplus F_3 \bigoplus F_3 \bigoplus F_1$.

\subsection{Simultaneous eigenstates of parity and the shift operator}

The vectors $|\tilde{\phi}^k_l\kt$ are however not eigenstates of parity $R$. In order to
construct this we find an unitary operator $H$ of the form $\otimes^K {\cal H}$, so that it 
commutes with $S$, and such that ${\cal H}$ diagonalizes ${\cal R}$. This fixes 
\beq
{\cal H }= \frac{1}{\sqrt{2}} \left( \begin{array}{rr} 1& 1 \\1 & -1 \end{array} \right),
\eeq
and $H$ as the Walsh-Hadamard transform. It follows that 
\beq
R H = H t, \;\;\; t=\mbox{diag}(1,-1,-1,1,\ldots),
\eeq 
$t$ is a diagonal matrix whose entries are the Thue-Morse sequence \cite{Allpaper}.
 Its $n$-th term is $t_n=(-1)^r$, where $r=\sum_j a_j$ and $a_j$ are the bits in the binary 
expansion of $n$. They satisfy the iterative rule:
\beq
\label{thuerule}
t_{2n}=t_n, \;\; t_{2n+1}=-t_n, \;\; t_0=1.
\eeq
Stated otherwise, the columns of the Walsh-Hadamard matrix
$H$ have parities that are arranged according to the Thue-Morse sequence. 

Consider the orthonormal complete set
 $|\phi^{k} _l \kt \, =\, H |\tilde{\phi}_l ^{k} \kt$. Since $S$ and $H$ commute, this is 
 clearly an eigenstate of the shift $S$. That it is also a parity eigenstate follows from:
 \beq
 R |\phi^k_l \kt =
    \f{1}{\sqrt{d}} \sum_{m=0}^{d-1}  e^{-2 \pi i lm/d}\, S^m R H  |k \kt\, =
     \f{1}{\sqrt{d}} \sum_{m=0}^{d-1}  e^{-2 \pi i lm/d}\, S^m H t |k \kt\, =
     t_k |\phi^k_l \kt.
 \eeq
 where we have used that $R$ commutes with $S$ and $t_k$ is the $k^{th}$ member
 of the Thue-Morse sequence. Thus:
  \beq
  \label{phi2}
  S |\phi_l ^{k} \kt = e^{2 \pi i l/d} |\phi_l ^{k} \kt, \;\; R |\phi_l^k \kt = t_k |\phi_l^k\kt, 
\eeq
We will also presently adapt these eigenstates to ``time-reversal'', however before that we
 notice that the plain Walsh-Hadamard transform shares some  common rows with
  $\br \phi^{k}_l|n \kt $. For instance when
 $k=0$ ($\sg=(0)$) and when $k=N-1$ ($\sg=(1)$) the rows 
 $\br \phi^{0}_0|n\kt$ and $\br \phi^{N-1}_0|n\kt$ consisting of all ones, and the
 row with the Thue-Morse sequence respectively are common to the Hadamard matrix.
  Indeed since the Thue-Morse and closely allied sequence dominate the eigenfunctions of the
  quantum baker's map, the Hadamard transform works well in this context.

 To be more explicit about the proposed transform we again illustrate with the case $K=3$.
 Using the ordering described above, the structure of the transform is 
 
 \beq
\dfrac{1}{\sqrt{8}} \left( \begin{array}{rrrrrrrr}
 1& 1& 1& 1& 1& 1& 1& 1\\
 1&-1& 1& 1&-1& 1&-1&-1\\ 
 1& 1&-1& 1&-1&-1& 1&-1\\
 1&-1&-1& 1& 1&-1&-1& 1\\
 1& 1& 1&-1& 1&-1&-1&-1\\ 
 1&-1& 1&-1&-1&-1& 1& 1\\
 1& 1&-1&-1&-1& 1&-1& 1\\
 1&-1&-1&-1& 1& 1& 1&-1\\
 \end{array} \right)\; \cdot \; \dfrac{1}{\sqrt{3}}
 \left( \begin{array}{cccccccc}
 \sqrt{3} & & & & & & & \\
  & 1&1 &1 & & & & \\
  & 1& \om & \om^2& & & &  \\
 &1 & \om^2& \om & & & &  \\
 & & & & 1& 1& 1&  \\
 & & & & 1&\om & \om^2& \\
 & & & & 1 & \om^2& \om&  \\
 & & & & & & &  \sqrt{3} 
 \end{array} \right),
 \eeq
where $\om=e^{2 \pi i/3}$. The second matrix is what we called $T_f$, 
the first is essentially the Walsh-Hadamard matrix, but for
the rearrangement of the columns according to cycles. Thus
this transform which appears to be new, combines the DFT
 and the Hadamard transforms in an interesting way.
 
\subsection{Time-reversal adapted states of the shift operator}
 
 The time-reversal symmetry of the quantum baker's map \cite{Qbmap} is not the same as that 
 of the shift $S$ as $G_N S G_N^{-1} \neq S^{-1}$. Therefore we cannot have the eigenstates
 of $S$ to be such that its Fourier transform is identical to its complex conjugate.
 However there is an analogous symmetry: 
 \beq
 {\mathsf b}_0 S {\mathsf b}_0= S^{-1},\;\; {\mathsf b}_0\, |a_{K-1}a_{K-2}\cdots a_{0}\kt = |a_{0}a_{1} \cdots a_{K-1}\kt.
  \eeq
 Here ${\mathsf b}_0$ is the bit-reversal operator, it reverses the significance 
 of the bits in a binary string. Clearly ${\mathsf b}_0^2 =1$. Its emergence
 is linked to the fact that the periodic points of the baker's map, are such that the
 position and momentum are bit-reversals of each other. Its connection to the Fourier
  transform is made even more closer in the context of the baker's map when we note that 
 \beq
 B=(G_N^{-1})_0 \, (\mathds{1} \otimes (G_{N/2})_1)
 \eeq
where the additional subscripts on the Fourier transform refer to the number of most
 significant bits that are left out while performing the transform, which is therefore a
 "partial Fourier transform" as defined by \cite{SchackCaves}. In particular $(G_N)_0$ is
 the full transform all the qubits and is therefore of dimensionality $N=2^K$, while
 $(G_{N/2})_1$ transforms the $K-1$ least significant bits and has the dimensionality 
 $N/2=2^{K-1}$, the first qubit is left unaltered. This is identical to the usual quantum 
 bakers map in Eq. (\ref{bvsbak}). Analogously it is easy to see by acting on bit strings
 that 
 \beq
 S=\mathsf{b}_{0}\, (\mathds{1} \otimes \mathsf{b}_{1})
 \eeq
 where $\mathsf{b}_k$ bit reverses the $L-k$ least insignificant bits,
  for example $\mathsf{b}_0$ reverses the whole string.

 The time-reversal symmetry of $S$ then implies that eigenstates $|\psi\kt$ of $S$ may be chosen such that 
 $\mathsf{b}_0|\psi\kt  = |\psi^*\kt$, where the complex conjugation is done in the standard
 position basis. Now the state $|\phi^k _l\kt$ need not be of this kind, therefore
 it is necessary to multiply by suitable phases or take appropriate linear combinations of them.
 Towards this end we define two bit strings $\sg$ and $\sg'$ shift-equivalent ($\sg \sim \sg')$
 if one is the result of repeatedly applying the cyclic shift operator $S$ to the other.
  There are two kinds of binary strings, we label them type-A and type-B. 
 If $\sg$ is a string of type-A then $\mathsf{b}_0 \sg \sim \sg$, and it is otherwise of
  type-B. It is somewhat surprising that the smallest binary string of type-B is of 
  length $6$ and that there are only $2$ of them: $ 110010,110100$, apart from
  their cyclic shifts. It is easy to see that if $\sg$ is of type-B then $S^m\sg$ is also
  of type-B for any integer $m$.
   
 If $\sg$ (value $k$) is a string of type-A then there exists an integer $p$ such that
  $S^p \sg = \mathsf{b}_0 \sg$. Also we note that $\mathsf{b}_0$ commutes with the
   Hadamard matrix $H$. Using these facts, a short 
 calculation shows that if
 \beq
 |\psi^{k}_l\kt = e^{i \pi l p/d} |\phi^{k}_l \kt, \;\;\mbox{then}\;\;
  \mathsf{b}_0|\psi^{k}_l\kt = |\psi^{k\, *}_l\kt.
 \eeq
Thus type-A strings give rise to states that are time-reversal adapted up to
 multiplication by a phase. If $\sg$ is of type-B, and its bit-reversal $\mathsf{b}_0 \sg$ 
 evaluates to $k'$, we note that $\mathsf{b}_0|\phi^k_l\kt= |\phi^{k'\, *}_l\kt$ and 
  $\mathsf{b}_0|\phi^{k'}_l\kt= |\phi^{k\, *}_l\kt$. Therefore we can
   construct the linear combinations:
 \beq
|\psi^{k\pm}_l \kt = \df{e^{-i \alpha_{\pm}}}{\sqrt{2}}
\left(|\phi^{k}_l\kt \pm  |\phi^{k'}_l\kt \right), 
 \eeq   
 where $\alpha_{+}=0$ and $\alpha_{-}=\pi/2$, which are orthogonal and such that 
 $\mathsf{b}_0|\psi^{k \, \pm}_l \kt = |\psi^{k \, \pm \, *}_l \kt$.
Thus this way we can construct a transform whose elements are $\br \psi_j|n\kt$, where $j$
labels all the parity and time-reversal adapted states of the shift-operator $S$.
In practice we order binary strings in the following way:
 $0,1,10,100,110,1000,1100,1110,10000,10100,\ldots,$
  such that if $\sg$ and $\sg'$ are two members on the list then
   $\sg$ or ${\mathsf b}_0 \sg$
are not shift-equivalent to $\sg'$. Among the possible representatives we choose
 that which is {\it largest}
when treated as a binary representation of an integer. Choosing the 
largest, as opposed to the smallest as in the previous case, gives us
 an unique increasing sequence that appears to be new. Given any $N=2^K$, we choose from this list strings whose length are
divisors of $K$. These would include strings of both type-A and type-B, in either case we construct 
a set of states based on the algorithm outlined above, finding $p$ by inspection in the 
case of type-A strings.     

\begin{figure}
\includegraphics[width=6in]{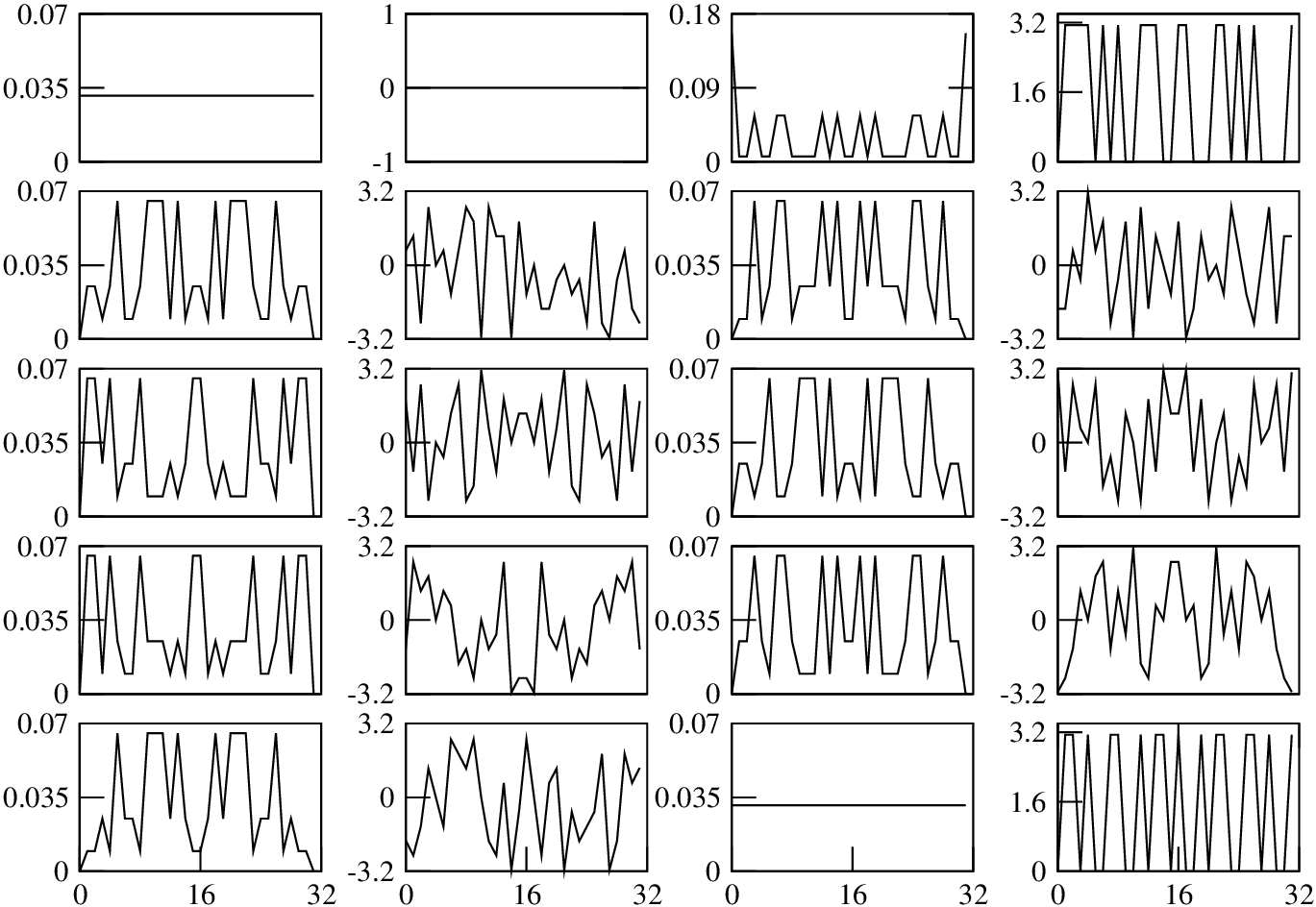}
\caption{Ten basis states $|\psi\kt$ that are simultaneous eigenfunctions of the shift operator, parity
(bit-flip) and time-reversal (bit-reversal) for the case $N=32$. Shown in the first and
third columns are the intensities ($|\br n|\psi\kt|^2$) while the second and fourth
columns have the corresponding phases. The first state is the uniform state, while the last 
has the Thue-Morse sequence as the components.} 
\label{basis3}
\end{figure}

In Fig.~(\ref{basis3}) we show ten of these states for the case $N=32$, showing the
simplest uniform state, the Thue-Morse sequence and eight other states that involve 
appropriate combinations of Fourier transforms of the columns of the Hadamard matrix.

\section{The eigenstates of the quantum baker's map in the new basis}

If $|\Phi_k\kt$ is an eigenvector of $B$ then we refer to its representation in the 
position basis as basis-0 ($\br n|\Phi_k\kt$), in the Hadamard basis as 
basis-1 ($\br n |H|\Phi_k\kt$), in the parity adapted basis of the shift operator
 as basis-2 ($\br \phi^{\sg}_l|\Phi_k\kt$), in the parity and time-reversal adapted basis
of the shift operator as basis-3 ($\br \psi_j|\Phi_k\kt$). 

\begin{figure}
\includegraphics[width=6in]{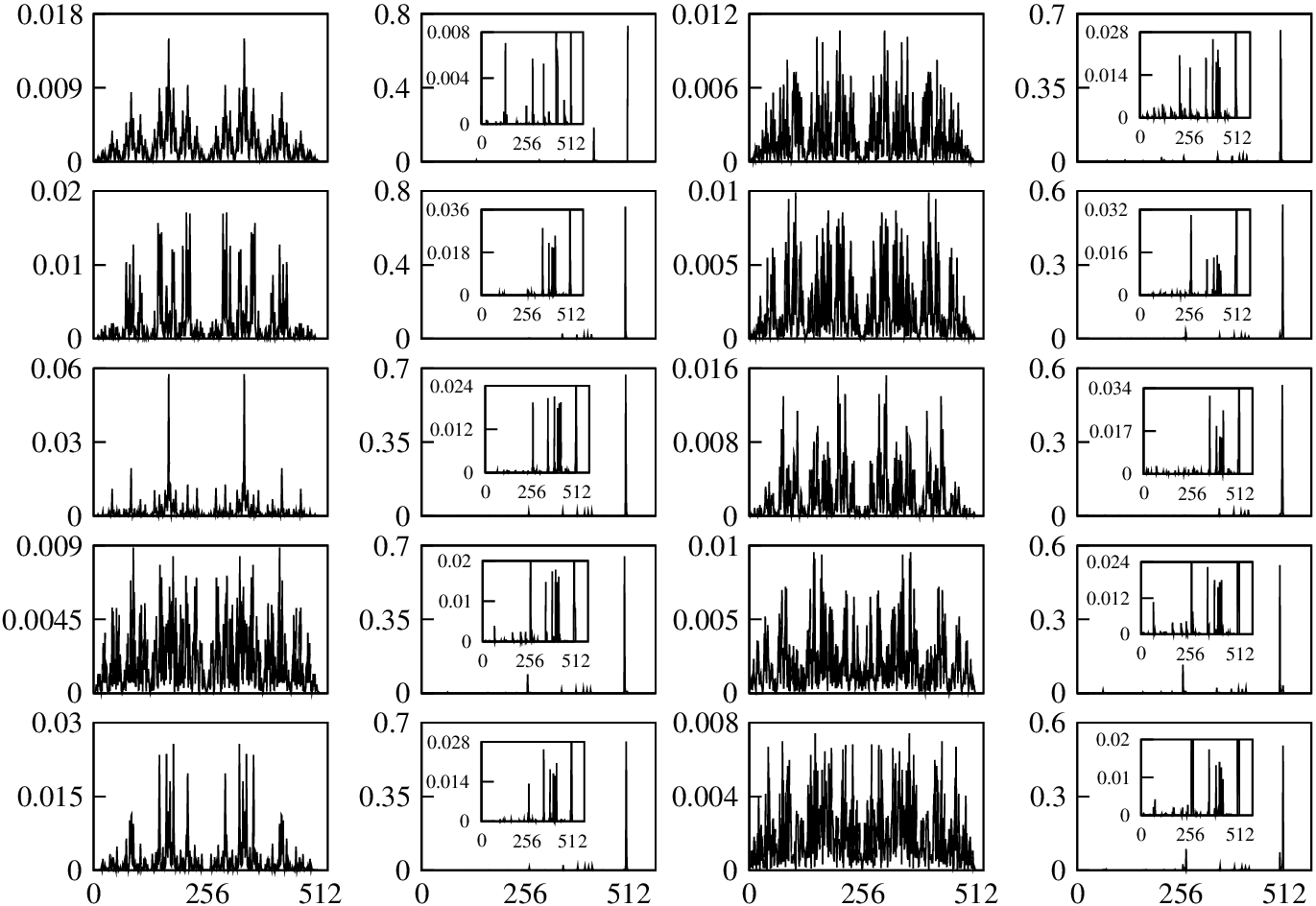}     
\caption{Ten eigenstates of the quantum baker's map for $N=512$ that are most simplified 
on using basis-2 or basis-3. Shown in columns 1 and 3 are the intensities of the states
 in the position basis. Columns 2 and 4 show the corresponding intensities on using
  basis-2 (along with the ``grass'' in the insets), this basis having states that are
   simultaneous eigenstates of the shift operator and parity. The Thue-Morse state is the
   first one.} 
\label{efs}
\end{figure}

In Fig.~(\ref{efs}) we show 10 most simplified eigenfunctions for $N=2^9=512$ using the new representations.
For comparison we show the states in both basis-0 (position) and in basis-2. We found that 
for these states basis-3 did not yield further significant simplifications, we will presently
quantify the degree of simplification across the spectrum. The state that is most simplified
 continues to be the Thue-Morse state which we have discussed in an earlier work. Its transformation
 whether using the Hadamard or the new transformations continues to be dominated by the 
 Thue-Morse sequence $t_k$. As we have pointed out earlier this state is well described 
 by the ansatz $t_K + G_N\, t_K$ \cite{meen}, where $t_K$ is the normalized Thue-Morse
  sequence of length $2^K=N$
 treated as a vector. In other words the Thue-Morse sequence and its Fourier transform dominates the
 state. The Fourier transform of the Thue-Morse sequence is known to be a multifractal \cite{Luck} 
 and we established that the quantum eigenstate we called the Thue-Morse state was also 
 a multifractal \cite{meen}. The structure of the eigenstate is dominated by the Fourier transform 
 of the Thue-Morse sequence, the peaks being much larger than $1/\sqrt{N}$. The peaks occur
 at the period~-~2 orbit of the doubling map (at 1/3,2/3) and at points corresponding to 
 homoclinic orbits of this point. Since the Thue-Morse states form a sequence for 
 increasing $N$, and $N\rightarrow \infty$ is the classical limit, it maybe expected that the
 Thue-Morse state is related to some classical invariant measure of the classical baker's map.
 We now show that this is indeed the case.
 
 Since we are dealing with the position representation of the states, and since the baker's
 map is such that its position coordinate evolves (in the forward direction) 
 independently of the momentum according to the doubling map $(x \, \mapsto 2 x \, (\mbox{mod}\, 1))$, the invariant density 
 is simply that of this one-dimensional map.
  If $\rho(x)$ is an invariant density of this map, we must have that:
 \beq
 \rho(x)= \f{1}{2}\,\rho\left(\f{x}{2}\right)\, +\, \f{1}{2} \,\rho\left(\f{x+1}{2}\right).
 \eeq
If $\rho(x)=\sum_{k=0}^{\infty} t_k \, \exp(2 \pi i k x)$ then it is easy to see, using the
recursion relation we have stated earlier namely $t_{2k}=t_k$, that it satisfies the
 requirement of an invariant density. Since the Thue-Morse state is close to the Fourier
 transform of the Thue-Morse sequence, it is suggestive that $\rho(x)$ is relevant to the
  classical limit of such states. Thus we believe we have a
  concrete example of a set of states of a quantum chaotic system that limits to a classical
   invariant measure that is not
 the ergodic measure, which would be uniform in this case. Instead it is a multifractal measure
 that is strongly peaked at period~-~2 periodic orbits and all their homoclinic excursions. 
 This is opposed to known examples where the limit is either ergodic or have delta-peaks
 corresponding to classical periodic orbits \cite{Nonnencat}. While we have given evidence
  of this without establishing it rigorously, this is an interesting deviation from quantum ergodicity
  that is allowed by Schnirelman's theorem \cite{Schnil}.
 It is very likely that many other eigenstates of the quantum baker's map are also of this 
 kind, limiting to classical invariant measures that are non-ergodic and are multifractal.
 That the eigenfunctions are multifractal we have already indicated in an earlier work \cite{meen}.
 
 It is pertinent here to connect with the works of Nonennmacher and co-workers, who have studied
 the quantum cat maps \cite{Nonnencat} and the Walsh-quantized baker's map \cite{nalini},
  which is an exactly solvable toy model of the bakers map. Among other things they have found two types of states in the
 Walsh-quantized bakers map, the first which they call ``half-scarred" has in the semiclassical
 limit part of its measure on classical periodic orbits and part is equidistributed in
 the Lebesgue measure. Such states have also been constructed by them for the quantum cat maps \cite{Nonnencat}.
 The tensor-product states found for the Walsh-quantized bakers map \cite{nalini} have semiclassical measures
 that are singular Bernoulli measures and were constructed as tensor products of states of
 the underlying ``qubit" space (when $N$ is a power of $2$ as in this paper, but have been generalized
 to other powers). The Thue-Morse states of the quantum bakers map (``Weyl" quantized) under
 discussion seem to be closer to the tensor-product states than the "half-scarred" ones. For one,
 there is strong evidence that the measure is multifractal in the semiclassical limit, and also
 the scarring can be unambiguosly associated with short period periodic-orbits and homoclinic ones \cite{meen}. 
 Moreover the Thue-Morse state is "close" to the simple tensor product $\otimes^K(|0\kt -|1\kt)/\sqrt{2}$
for $N=2^K$, which indeed is the finite Thue-Morse sequence. This can be measured in terms of the modulus of the inner-product and while this
does decrease with $N$, it does so slowly. Numerical calculations not shown here indicate a decay of
 $N^{-0.1}$. Note that typical inner-products with random states will scale as $N^{-0.5}$.
 A more accurate representation of the Thue-Morse state as a superposition of the above product and its Fourier transform is such 
that their inner product decays even more slowly to zero (as $N^{-0.08}$). We postpone a more detailed description
to a later publication, suffice to say that the resolution of the semiclassical measure of the
 Thue-Morse state and in general other states of the quantum bakers map into
  singular-continuous, pure-point and continuous components is not yet clear.

To quantify the extent to which the eigenvectors $|\Phi_k\kt $ of $B$ are simplified
 we evaluate the participation ratios (PR). When a complete orthonormal basis
 $ \{|\alpha_i\kt, \, i=0,\ldots,N-1\}$ is used the PR is defined as
\beq 
\left(\sum_{i=0}^{N-1}\left|\br \alpha_i |\Phi_k \kt\right|^4 \right)^{-1}
\eeq
 The PR is an estimate of the number of $|\alpha \kt$ basis states needed to construct the vector 
$|\Phi_k\kt$, here chosen to be one of the eigenstates of $B$. We calculate the PR in 
(1) the position basis (basis-0, $|\alpha\kt =|n\kt$), (2) the Hadamard basis (basis-1, $|\alpha \kt=
H|n \kt$), (3) the basis that consists of parity reduced eigenstates of $S$ (basis-2,
$ |\alpha\kt = |\phi^k_l\kt$), and (4) the basis that has both  parity and time-reversal
 symmetry reduced eigenstates of $S$ (basis-3, $|\alpha \kt = |\psi_j \kt$).

\begin{figure}
\includegraphics[width=5in]{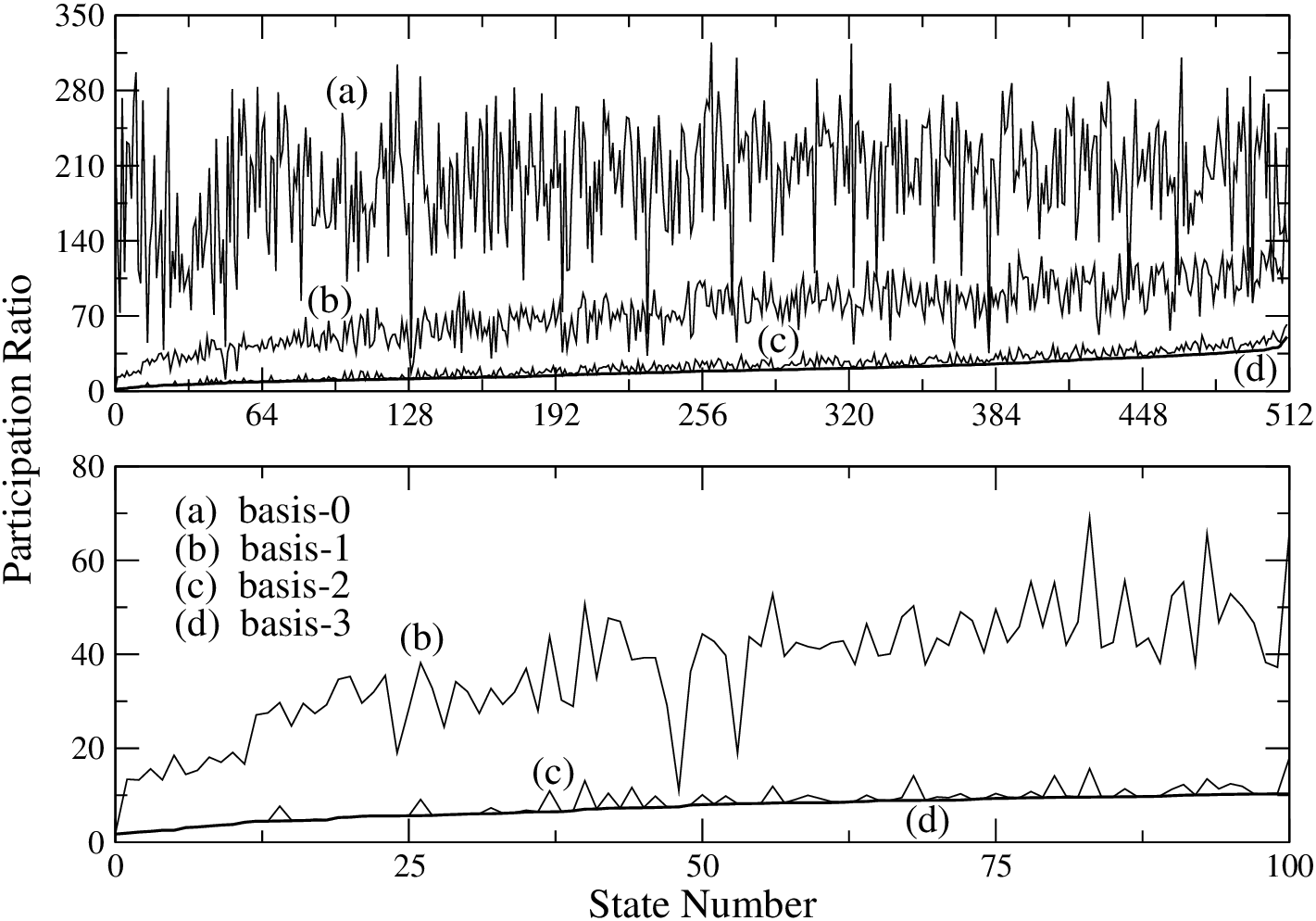}
\caption{The participation ratio of the eigenstates of the baker's 
map in the position basis (basis-0), the Walsh-Hadamard basis (basis-1), the
basis that is parity reduced eigenstates of $S$ (basis-2), and
in the basis that is also time-reversal symmetry adapted (basis-3).
The states are arranged in the increasing order of the PR in basis-3.
This is for the case $N=512$, and the lower figure is a magnification of details for
the first hundred states.}
\label{PR}
\end{figure}
From Fig.~(\ref{PR}) we indeed see that the eigenstates of $S$, properly symmetry 
reduced simplify the eigenstates of the quantum baker's map $B$ significantly.
This does considerably better than the previously used Walsh-Hadamard transform. 
The difference is only however marginal for the Thue-Morse state, as the last column
of the Walsh-Hadamard transform, the Thue-Morse sequence of finite generation,
is anyway a parity and time-reversal symmetric eigenstate of the shift operator
$S$. We see from the figure that for $N=512$, there are about hundred
states that can be constructed from ten or fewer eigenstates of the shift
operator as we have constructed them above. We also notice that while there is
considerable difference between using the basis-2 and the Hadamard basis, basis-1,
there is not that much simplification due to the use of basis-3 over basis-2. This
is understandable as parity symmetry ($R$) is the same for both $S$ and $B$, while the
time-reversal symmetries are different. 

Due to the simplicity and efficacy of basis-2, we will discuss this further. 
We first point out that the dual or momentum  basis $G_N|\phi_l^{k}\kt$ is exactly 
as effective as the original basis for studying the eigenfunctions of the 
baker's map, due to time-reversal symmetry. To prove this notice that 
\beq
G_N|\phi^k_l\kt = -G_N^{-1}\, R \, |\phi^k_l\kt=-t_k G_N^{-1}\, |\phi^k_l \kt
\eeq
where we have used $G_N^2=-R$ (for example see Saraceno in \cite{Qbmap}), and Eq.~(\ref{phi2}).
Using time-reversal of the eigenstates of $B$ $(G_N |\Phi\kt =|\Phi^*\kt)$ implies that 
\beq
\br \Phi |G_N^{-1}|\phi_l^{k} \kt = \left(\sum_n \br \Phi|n\kt \br n |\phi^{k}_l \kt ^*
\right)^*=\br \Phi |\phi^{k}_{l'}\kt^*
\eeq
where $l'=d-l$ unless $l=0$, in which case $l'=0$ as well. Hence finally
\beq
\br \Phi|G_N |\phi^k_l \kt = -t_k \br \Phi |\phi^{k}_{l'} \kt^{*}.
\eeq
Thus while the overlaps of the eigenstates of the quantum baker's map in a basis which is the  the 
Fourier transform of the basis-2 are not the same as originally, they are upto a sign, complex conjugates
of overlaps with some other basis states with a different value of $l$ in general. Clearly this 
leads to identical participation ratios. Thus the Fourier transform of basis-2
is also of interest, indeed as we have already indicated, many of the eigenstates of
 the baker's map look like them, the foremost being the Thue-Morse state where
  the momentum representation of $|\phi^{(N-1)}_0\kt$ is of relevance. Thus we have a natural way of
generalizing this class of functions. We do not pursue this further here,
 but note that the other Fourier transforms are also of relevance to the spectrum of the
 quantum baker's map, and they also have multifractal characters. A few related functions
 have recently been studied by us as the "Fourier transform of the Hadamard transform"
 \cite{NCNSD}. It is also reasonable to expect that a combination of basis-2 and its Fourier
 transform maybe even closer to the actual eigenstates of the quantum baker's map than
 even basis-2. For instance in the case of the Thue-Morse state, such a combination 
 does better than either the Thue-Morse sequence or its Fourier transform 
 taken individually \cite{meen}. 
 
 To summarize, in this work we have constructed eigenfunctions of the shift operator that 
 have additional symmetries of bit-flip or parity and bit-reversal or time-reversal. Using these
 we have seen why the Walsh-Hadamard transform simplifies states of the quantum baker's map, as
 well as shown that these transforms are capable of doing significantly better. The use of these
 transforms in other contexts, other than the quantum baker's map, is possible. It combines
 elements of both the Fourier and the Hadamard transforms in an interesting way. Using these
 transforms helps us study the eigenfunctions of the baker's map in a more detailed manner, and 
 our future work will explore this further. Operators akin to the shift operator have been used as
 toy models of open quantum bakers to study fractal Weyl laws \cite{Nonnen}. It is also 
 interesting that the same cocktail of the shift operator, the Fourier transform and the
 Hadamard transform appears essentially in Shor's quantum algorithm for factoring, a fact 
 also previously pointed out in \cite{aruljphys}. A very recent work makes a significant
  contribution by constructing suitable basis sets for $N$ that are not powers of $2$
  \cite{saraceno06}.

\acknowledgments{A.L. thanks Andr\'e Voros and Marcos Saraceno for 
discussions and interest regarding this and related work.
 He is grateful to St\'ephane Nonnenmacher for conversations and e-mails
clarifying the action of the shift operator and the role of Hecke eigenfunctions
in the quantum cat maps that led to seeking symmetry reduced states used in this paper.
 N.M. was supported by financial assistance from the Council for Scientific and
 Industrial Research, India.}

\end{document}